\documentclass[12pt,english]{article}

\makeatletter

\providecommand{\tabularnewline}{\\}

\usepackage{amsmath,amssymb}
\usepackage{graphicx}
\usepackage{cite,floatflt,here}

\newcommand{\PPrtNo}
{arXiv:0903.2667 [hep-ph] \hfill SMU-HEP-09-05 \\}
\newcommand{\TITLE}
{Improved formulation of global QCD analysis\\ with zero-mass hard cross sections}

\newcommand{\AUTHORS}
{Pavel M.~Nadolsky$^{a}$, Wu-Ki Tung$^{b,c}$}
\newcommand{\INST}
{
$^a$ Department of Physics, Southern Methodist University, Dallas, TX 75275, USA\\
$^b$ Department of Physics and Astronomy, Michigan State University,\\ East Lansing, MI 48824, USA\\
$^c$  Department of Physics, University of Washington, Seattle, WA 98105, USA }

\newcommand{\ABSTRACT}
{The zero-mass (ZM) parton formalism is widely used in high-energy
physics because of its simplicity and historical importance, even while
massive quarks ($c,b,t$) are playing an increasingly prominent role in
particle phenomenology, including global QCD analyses of parton
distributions based on the more precise general-mass (GM) QCD
formalism. In view of this dichotomy, we show how the obvious
inconsistencies of the conventional implementation of the ZM formalism
can be corrected, while preserving the simplicity of its hard matrix
elements. The resulting \emph{intermediate-mass} (IM) scheme for
perturbative QCD calculation can be considered either as \emph{improved
ZM formulation} with realistic treatment of heavy-flavor kinematics; or
as a \emph{simplified GM formulation} with approximate ZM hard cross
sections. Phenomenologically, global analyses based on IM calculations
can effectively reproduce, within the present estimated uncertainty
bands, the more correct GM results on parton distributions, as well as
their predictions for a wide range of collider processes of current
interest. \\ \quad\\ PACS: 12.15.Ji, 12.38 Cy, 13.85.Qk }

\makeatother

\usepackage{babel}

\begin{document}
\begin{titlepage}

\begin{tabular}{l}
\tabularnewline
\end{tabular}

\noindent \PPrtNo

\vspace{1cm}

\begin{center}
\renewcommand{\thefootnote}{\fnsymbol{footnote}}  {\large \TITLE
{}}
\par\end{center}{\large \par}

\begin{center}
\vspace{1.25cm} {\large \AUTHORS{}}
\par\end{center}{\large \par}

\begin{center}
\vspace{1.25cm}
\par\end{center}

\begin{center}
\INST

\par\end{center}

\vfill

\ABSTRACT\vfill

\newpage
\end{titlepage}

\tableofcontents{}\newpage



\quad\\

\section{Introduction}

Global QCD analysis is based on factorization theorems of perturbative
quantum chromodynamics (PQCD) at high energies. It allows the determination
of universal parton distribution functions (PDFs) by comparing the QCD
parton formulas with a wide range of available hard scattering data (which
involve at least one large energy scale, generically referred to as $Q$).
The conventional derivation of the factorization theorems in PQCD is
formulated in the zero-quark-mass limit and is valid to $\mathcal{O}$($%
\Lambda _{QCD}^{2}/Q^{2}$) \cite{ZMfac}. Since QCD interactions depend on
the number of active quark flavors, and this number (denoted by $n_{f}$)
varies in practice, it is necessary to vary $n_{f}$ in the theoretical
calculations according to the effective energy scale $\mu $ of the
interaction---it is incremented by $1$ every time $\mu $ crosses one of the
mass thresholds for heavy quarks \cite{BernreutherMarciano}. In the common
implementation of this so-called variable-flavor-number-scheme (VFNS), the
changes of $n_{f}$ with the factorization scale across heavy-flavor
thresholds are taken into account in the evolution of PDFs and in the
summation over active parton flavors in the factorization formula, but 
heavy-quark mass effects are ignored in the hard matrix elements, as well as in the
evaluation of the phase space of final states in the convolution integral.
This provides a simple and convenient working platform for most calculations
in high-energy phenomenology; and it has been widely used both in practical
calculations and in global QCD analysis of PDFs. We shall refer to this
calculational scheme as the \emph{conventional} zero-mass (ZM) VFNS.

As the PQCD theory advanced, and the proper treatment of heavy quarks became
more important, it was recognized that the validity of factorization ought
to be extendable to PQCD with non-zero quark masses \cite{ColTun,ACOT},
\emph{i.e.} to be valid to order $\Lambda _{QCD}^{2}/Q^{2}$ uniformly,
independent of the ratios $m_{h}/Q$ of quark masses $m_{h}$ for $h=c,$ $b,$
and $t$. This was formally proved for inclusive deep inelastic scattering in
Ref.~\cite{collins}. This generalized factorization, which does not assume $%
m_{i}/Q\rightarrow 0$ for all parton flavors $i$ as in the ZM formalism,
provides the basis for an improved global QCD analysis formulated in the
general-mass variable-flavor-number scheme (GM VFNS). Recent GM VFNS global
analyses \cite{cteq65,cteq66,MSTW08} demonstrate that the proper treatment
of heavy-quark mass terms is essential for reliable predictions of cross
sections at the Tevatron and LHC. This is because a good fraction of input
precision data from DIS and fixed-target experiments included in the global
analysis are at energy scales comparable to, or not too far above, the charm
and bottom masses \cite{ThorneTung}; hence the PDFs are sensitive to the
more precise treatment of mass effects in the GM scheme.

Nonetheless, because the ZM PQCD formalism is much better known, and the
requisite hard cross sections (Wilson coefficients) are much more readily
available, the ZM VFNS remains ubiquitously used in particle physics
applications, as well as in many contemporary global analyses of parton
distributions \cite{cteq60,cteq61,ZMfits}. With this in mind, it is useful
to reexamine the ZM calculations from a new perspective, to see whether it
is possible to preserve its main simplifying feature, massless
hard-scattering cross sections,\footnote{%
It should be noted that ZM and GM hard cross sections are practically
indistinguishable if the typical energy is much larger than $m_{c}$ and $%
m_{b}$, as is the case in most collider phenomenology. In this
high-energy regime, ZM hard cross sections are compatible with the
PDFs obtained by global analyses in either mass scheme. What we are
concerned here are the differences between the ZM and GM calculations
at low $Q$ that affect the global analysis of the precise DIS data from
HERA at small $x$. The resulting differences in PDFs can affect 
predictions of physical observables at both low and high
energies.\label{footnoteZMGM}} while correcting its other obvious
inconsistencies. These inconsistencies, largely overlooked before, were
recognized in \cite{cteq65}; the possibility of formulating improved
implementation of the ZM scheme was raised in \cite{ThorneTung}. In
this paper, we address this problem in detail. We are motivated by the
observation that the differences between the GM and ZM PDFs that caused
the observed shifts in the LHC $W$ and $Z$ cross section predictions
arise mainly due to the kinematical suppression of heavy-flavor
scattering cross sections near their respective mass thresholds in the
global analysis. Inclusion of the correct kinematic treatment in the ZM
global analysis can bring the resulting PDF's much closer to the GM
analysis, while retaining the simplicity of the massless hard cross sections 
in the ZM scheme, if this is done with due care. We focus on
the next-to-leading order (NLO) global analysis, which 
benefits most directly from such improvements. 

In Section \ref{sec:conven}, we analyze the essential components of the
variable flavor number scheme and identify those parts of the
conventional ZM implementation that conflict with kinematic aspects of
heavy-flavor production. Based on this analysis, in
Section~\ref{sec:alt}, we propose practical methods to correct these
problems, leading to a class of ``intermediate-mass" (IM) schemes that
potentially can serve as effective approximations to the GM scheme.
Numerical comparisons of ZM/IM/GM calculations are presented. In
Section~\ref{sec:GlbFits}, we perform parallel global fits based on the
ZM/IM/GM schemes, compare the effectiveness of the proposed IM schemes,
and compare the PDFs and theoretical predictions for representative
collider cross sections. Alternative choices of the rescaling variable
in the GM scheme are considered in Section~\ref{sub:zetaGM}. 
Concluding remarks are given in Section~\ref{sec:Conclusions}.

\section{Conventional ZM VFNS\label{sec:conven}}

The general PQCD factorization for high-energy hard processes, exemplified
by the inclusive DIS structure functions $F_{\lambda }(x,Q^{2})$, has the
form
\begin{eqnarray}
F_{\lambda }(x,Q^{2}) &=&\sum_{a,b}\int_{\chi }^{1}{\frac{d\xi }{\xi }}%
\,\,f_{a}(\xi ,\mu )\,C_{b,\lambda }^{a}\left( \frac{\chi }{\xi },\frac{Q}{%
\mu },\frac{m_{i}}{\mu },\alpha _{s}(\mu )\right)  \notag \\
&\equiv &\sum_{a,b}\left[ f_{a}\otimes C_{b,\lambda }^{a}\right] (\chi
,Q,m_{i},\mu ).  \label{master}
\end{eqnarray}%
Here $f_{a}(\xi ,\mu )$ is the parton distribution function for an
initial-state parton $a$ with momentum fraction $\xi $ at the
factorization scale $\mu $, $C_{b,\lambda }^{a}$ are hard cross
sections for the scattering of parton $a$ into a final-state parton
$b$, and $m_{i}$ collectively represents the quark masses. The
perturbatively calculable hard cross sections $C_{b,\lambda }^{a}$ are
also variously known as Wilson coefficients, hard matrix elements, or,
least formally, matrix
elements. In VFNS, initial states $a$ are summed over
the active parton flavors at the factorization scale $\mu $, typically
chosen to be $\mu =Q$ (and equated here to the renormalization scale).
The lower limit of the convolution integral $\chi $ is a function of
the DIS kinematic variables ($x,Q$) determined by final-state
phase-space constraints in each parton-level scattering subprocess.

In its general form, the factorization formula (\ref{master}) applies both
to the GM and ZM VFNS. With the insight gained from the modern formulation
of the GM VFNS, we will now dissect this formula in the conventional
implementation of the ZM formalism and expose the elements that are usually
taken for granted together. (Since we always work with the VFNS in this
paper, we shall omit the VFNS designation for both GM and ZM cases from now
on.) \medskip

\noindent (i) \textbf{ZM hard matrix elements}. 
Central to the ZM formalism, the
hard matrix elements in Eq.~(\ref{master}) are taken to be the standard ZM ones, $%
C_{b,\lambda }^{a}=C_{b,\lambda }^{a}\left( \frac{\chi }{\xi },\frac{Q}{\mu }%
,0,\alpha _{s}(\mu )\right) _{\overline{\mathrm{MS}}}$: they are calculated
under the assumption that all active partons are massless, with the
associated singularities subtracted in the $\overline{\mathrm{MS}}$ scheme.
Their expressions are well-known---to NNLO in the QCD coupling $\alpha
_{s}(\mu )$ for DIS and Drell-Yan processes \cite{NNLOZM}, and to NLO for
many other processes.

For the DIS process, the hard matrix elements are also known for non-zero
quark masses to order $\alpha_{s}^{2}$ \cite{NNLOGM}, but the zero-mass
formulas are much simpler and easier to use. Hence the ZM calculation
is commonly used \emph{because of its simplicity and convenience}. On
the other hand, for almost all other physical processes, such as
Drell-Yan pair, $W/Z$ boson, and inclusive jet production, the
GM hard matrix elements are not available beyond the leading
order. Thus, \emph{by necessity}, the ZM matrix elements are still
widely used both in physical applications and in global analyses in
general.

\medskip

\noindent (ii) \textbf{Final-state counting}. The summation of the
final-state parton flavors in Eq.~(\ref{master}), $\sum_{b}$, is
conventionally taken to be over all active parton flavors at scale $\mu$,
the same as for the initial-state summation $\sum_{a}$.

This convention originated from the strict zero-mass parton formalism
of the 1970's and 80's, but is clearly problematic from the modern
perspective. As emphasized in \cite{cteq65,ThorneTung}, there is an
important conceptual difference between the initial- and final-state
summations $\sum_{a}$ and $\sum_{b}$. $\sum_{a}$ runs over the active
parton flavors, a \emph{theoretical} concept dependent on the choice of
the renormalization scheme for $\alpha_{s}$, the factorization scheme
for the PDFs, and, of course, the value of $\mu$.
On the other hand, $\sum_{b}$ involves summation over \emph{%
physical} final states that are kinematically allowed at the given
scattering energy. Determined by  {\it physical} kinematical
considerations, $\sum_{b}$  should  be \emph{independent} of the choice of the
renormalization and factorization schemes and the scale variable $\mu$.

This dichotomy leads to unintended inconsistencies for the conventional
ZM calculation. For example, consider a small-$x$ kinematic
configuration common at HERA, say $x=10^{-4}$ and $Q=3$~GeV,
corresponding to virtual Compton scattering center-of-mass (CM) energy
$W=Q\:\sqrt{x^{-1}-1}=300$ GeV. Because the factorization scale
$\mu=Q=3$ GeV is smaller than the bottom quark mass $m_{b}$, the bottom
quark $b$ in this calculational scheme is not counted as an active
parton; thus the final-state parton flavor summation does not include
$b$. However, in reality bottom quarks are easily produced at this CM
energy $W$, as is indeed experimentally observed. The
problem is even more pronounced for $Q$ around and below the charm mass $%
m_{c}$.\medskip

\noindent (iii) \textbf{Phase-space treatment}. In ZM calculations, the
convolution integral $\int_{\chi }^{1}{\frac{d\xi }{\xi }}$ in Eq.~(\ref%
{master}) is usually computed using massless parton kinematics, so that $%
\chi $ in the lower integration limit is equated to the Bjorken $%
x=Q^{2}/(2q\cdot p)$. For the leading-order quark scattering process $%
V^{\ast }a\rightarrow b$ (where $V^{\ast }$ stands for the virtual vector
boson $\gamma ^{\ast }/W/Z$), this leads to the well-known result $%
F_{\lambda }(x,Q^{2})=\sum_{a,b}C_{b,\lambda }^{a}\,f_{a}(x,Q)$, where $%
C_{b,\lambda }^{a}$ are the appropriate (electroweak) coupling parameters.

Because this integral originates from summing over final-state phase space
of \emph{real} particles, this practice leads to violation of Lorentz
kinematics in the case of heavy-flavor production. For instance, in
neutral-current DIS at  $Q\gtrsim m_{b}$ and $W\lesssim2m_{b}$ 
(say, $x\sim0.3$), this
calculational scheme will predict similar contributions 
from $b$ quark production and $d$ sea quark
production (since ZM hard matrix elements are flavor-independent), whereas, in
fact, this kinematical regime is below the $b$-production threshold, and the
$b$ and $d$ scattering cross sections are completely 
different!\footnote{Additional kinematical effects may be relevant in semiinclusive observables, e.g., due to the heavy-quark masses in resummed multi-scale 
differential distributions \cite{ResumSIDIS} or final-state hadron masses in 
heavy-quark fragmentation \cite{KniehlEtAl}.}
\medskip

\noindent Past global QCD analyses carried out in the ZM scheme have
overlooked these unphysical features, since the heavy-quark
contributions to flavor-inclusive cross sections are relatively small.
The resulting deviations from the true behavior are compensated in the
global analysis process by the fitting process, leading to very good
agreement with the examined data. Thus, many ZM PDFs (including CTEQ6
\cite{cteq60} and CTEQ6.1 \cite{cteq61}) have been widely used
 in comparisons of current experimental measurements to
theory, and in
making predictions for new processes in collider phenomenology.\footnote{%
Parallel PDF sets in the GM VFNS, such as CTEQ5HQ/CTEQ6HQ
\cite{cteq5,cteq6HQ}, were also provided along with CTEQ6 and CTEQ6.1.
They have not been used as widely in general applications, but served
as standard PDFs in the analysis of heavy-quark production.}

Having identified the unphysical features of the conventional
implementation of the ZM scheme in this section, we can now investigate
whether these elements can be corrected, while preserving its simple
matrix elements, to produce a simple and effective approximation to the
GM theory. This is worthwhile, since the ZM formalism is very versatile
and practicable; and it is still used in the vast majority of applications.


\clearpage
\section{Intermediate mass VFNS\label{sec:alt}}

From the presentation of the last section, we can see that the
particulars of counting final states and of treating final state phase
space in the conventional implementation of the ZM scheme (points (ii)
and (iii) of the previous section) are not intrinsic to the ZM 
treatment of hard matrix elements (point (i)). With the insight gained
from the GM formulation of \cite{cteq65}, we can try to eliminate these
kinematic contradictions on inclusive cross section calculation from
the ZM scheme. There is no unique way to do this, even though the
underlying physics provides valuable guidance on how to proceed. We
shall now discuss the possibilities.

\subsection{Satisfying heavy-flavor production kinematics \label%
{sec:rescaling}}

First, consider the issue of \textbf{phase space integration} for a
given partonic subprocess contributing to the right-hand side of Eq.\,(\ref%
{master}), say the LO $V^{\ast }+q\rightarrow h$, where
$q$ denotes either a light or heavy quark, and $h$ a heavy quark ($%
c,b,t$). In the conventional implementation of ZM, for zero-mass
kinematics, one gets the contribution $F_{\lambda
}(x,Q^{2})=C_{h,\lambda }^{q}\,f_{q}(x,Q)$, which runs into serious
problems if $W$ is near or below the production threshold for the heavy
quark. It has been known since early days of charged-current DIS
\cite{Barnett}, and, more recently, from the GM approach
\cite{acotchi,cteq65,ThorneTung} that the phase space constraint due to
heavy-quark masses can be naturally implemented by replacing the
Bjorken $x$ above by a \emph{rescaling variable} ``ACOT-$\chi $",
\begin{equation}
\chi =x\left( 1+M_{f}^{2}/Q^{2}\right) ,  \label{acotchi}
\end{equation}%
where $M_{f}$ denotes the total mass of the final state ($m_{h}$ for
charged-current DIS, $2m_{h}$ for neutral-current DIS), \emph{i.e.} $%
\,F_{\lambda }(x,Q^{2})=C_{h,\lambda }^{q,(0)}\,f_{q}\left( \chi
(x,Q)\,,Q\right) $. One can easily verify that, as $W$ approaches the
heavy-quark production threshold $M_{f}$ from above, $\chi \rightarrow
1$; thus $F_{\lambda }\rightarrow 0$ in the heavy-quark production
channel, as
it should.\footnote{%
This rescaling is more than just a convenient prescription. Its
emergence can also be seen in the origin of the LO term as the
resummation of collinear singularities of higher order terms in the
limit $Q\gg M_{f}$: if mass effects are kept, this variable will appear
naturally in the collinear terms, cf.\,\cite{acotchi}.}

This problem is general, applicable also to NLO and beyond: by extending the
lower bound of the convolution integral to Bjorken $x$ in Eq.\,(\ref{master}%
), the conventional ZM formalism grossly overestimates the contributions
from the region of phase space near (and even beyond) the physical
thresholds \emph{for all channels}. A straightforward way to correct
this problem is to replace the lower limits of the convolution integral
with the equivalent rescaling variable $\chi(x,Q)$ for all terms. This
would restrict the phase space to the physically allowed region, and,
importantly, also ensure consistency between the LO and higher-order
terms with respect to a shift of the factorization scale. However, this
prescription is not unique; and it needs some careful consideration.
For instance, the rescaling prescription Eq.\,(\ref{acotchi}) involves
an upward shift of the variable $x$ by a constant factor $\left(
1+M_{f}^{2}/Q^{2}\right)$, even though the physical mass-threshold
kinematic constraint is mainly a large-$x$ (low-$W$) issue for a given
$Q$. Since parton distributions are rapidly varying functions at small
$x$, the suppression of phase space due to the substitution
$x\rightarrow \chi $ could unnecessarily suppress the contribution from
the gluon fusion term in the small-$x$ region, causing potential
disagreement with the accurate DIS data from HERA.

It is therefore desirable to generalize the rescaling variable $\chi$,
so that it enforces the mass-threshold constraint at large $x$ in the
same natural way, but smoothly recovers the standard scaling $x$
variable away from the mass threshold (\emph{i.e.}, at small $x$) in a
controllable way. We introduce a new rescaling variable $\zeta$ using
the relation, modeled after Eq.(\,\ref{acotchi}):
\begin{equation}
x=\zeta\left(1+\zeta^{\lambda}M_{f}^{2}/Q^{2}\right)^{-1},  \label{zeta}
\end{equation}
where the parameter $\lambda$ is a positive number. The variable $\zeta$ has
the following properties:\smallskip\newline
a) it reduces to Bjorken $x$ for large $Q$ ( $\zeta\rightarrow x$ as $Q\gg
M_{f}$), and the $\zeta\rightarrow1$ limit corresponds to the physical mass
threshold ($W\rightarrow M_{f}$)---both key features of a rescaling variable;%
\footnote{%
Specifically, one can show that $W^{2}=Q^{2}\left((1+\zeta^{%
\lambda}M_{f}^{2}/Q^{2})\zeta^{-1}-1\right)$, neglecting ordinary
hadron masses.}\smallskip\newline b) for non-zero $\lambda$, the
rescaling effect is reduced as $x$ moves toward smaller values, away
from the threshold region where it is required; and\smallskip\newline
c) for a given $Q$, $x<\zeta<\chi$, i.e.\,$\zeta$ lies in-between the
ZM $x$ (which violates non-zero mass kinematics) and the uniform
rescaling variable $\chi$ (which may be too restrictive).

\begin{floatingfigure}[r]{3in}
\includegraphics[width=2.6in]{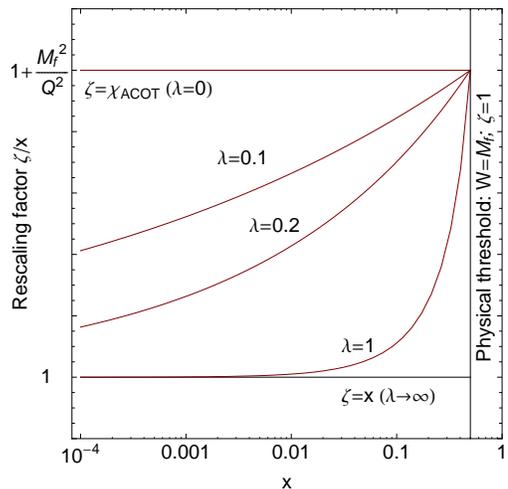} \caption{The
rescaling factor $\zeta/x$ vs. $x$ for $\lambda=0,$ $0.1,$ $0.2,$ and
$1$.} \label{fig:Zeta}
\end{floatingfigure}
The rescaling factor $\zeta/x$ is shown in Fig.\,\ref{fig:Zeta} for several
values of $\lambda$. The top horizontal line, $\lambda=0$, corresponds to
uniform rescaling $\zeta=\chi$. The base line corresponds to the ZM case $%
\zeta=x$. Those two lines are parallel, since $\chi/x$ is a constant factor,
as already mentioned. The other curves correspond to $\lambda=0.1,\,0.2$,
and $1$, respectively. The amount of suppression of phase space away from the
physical mass threshold can be controlled in a smooth way by varying the
parameter $\lambda$. \ The small-$x$ rescaling is reduced for increasing
values of $\lambda$; and it has practically vanished below $x=0.1$ even for
a moderate choice of $\lambda=1$.\footnote{%
The limiting case $\zeta=x$ is reached in Eq.(\ref{zeta}) formally by taking
$\lambda\rightarrow\infty$. But, as can be seen from the steep rise of the
rescaling factor curves for larger values of $\lambda$ in Fig.\,\ref%
{fig:Zeta}, the transformation as $x\rightarrow1$ is ill-defined in this
limit. For all practical purposes, $\lambda\approx1$ already produces enough
small-$x$ suppression of rescaling to produce results close to that using
directly $\zeta=x$, while ensuring kinematic bounds.} The range of variation
of the rescaling variable, represented by the width between the two
horizontal lines, is $M_{f}^{2}/Q^{2}$; it shrinks rapidly for increasing
values of $Q$. Thus, \emph{rescaling at high energy scales is negligible for
any choice of} $\lambda$\emph{\ for all} $x$.

The rescaling variable $\zeta$ is simple and flexible. It can be used
effectively to explore how well the conventional ZM formulation can be
improved to yield more reliable predictions. In the next section, we will
show examples of global fits based on improved ZM calculations with this
variable, with selected values of the parameter $\lambda$. We will
find that the generalized rescaling variable is the key 
requisite needed to correct the kinematical dependence of the ZM cross 
sections, while the other modifications (discussed in the next subsection) 
play a secondary role. 

\subsection{Correct summation over physical final states \label{sec:FS}}

Next, we turn to the issue of \textbf{summation over final-state partons}, $%
\sum_{b}$ of Eq.\,(\ref{master})---point (ii) raised in
Sec.\,\ref{sec:conven}. This is intimately related to the choice of
partonic subprocesses included in the calculation for the given
combination of kinematical parameters. Consider, in particular,
production of a $b$ quark in neutral-current DIS discussed in item (ii)
of the previous section ($x\sim10^{-4},$ $Q\sim3$ GeV, $W\sim300$ GeV).
Since $b\bar{b}$ pairs are physically produced in the final state, the
gluon fusion subprocess $\gamma^{\ast}g\rightarrow b\bar{b}$ must be
included in the sum of non-zero scattering contributions independently
of the choice of the factorization scheme and scale. In the
conventional implementation of the ZM scheme, with the scale $\mu=Q=3$
GeV being below the bottom mass, this
subprocess is not included, resulting in the dilemma mentioned in Sec.\,\ref%
{sec:conven}. But, there is no reason, in principle, to tie the final-state
parton summation $\sum_{b}$ to the initial-state one, $\sum_{a}$. In fact,
in the fixed-flavor-number scheme, as in the GM VFNS, the subprocesses $%
\gamma^{\ast}g\rightarrow c\bar{c},\, b\bar{b}$ are always present whenever
there is enough CM energy to produce these heavy-flavor states. Therefore,
we can try to adopt the same physically sensible approach in the ZM scheme,
i.e.\,we shall include these subprocesses while keeping the ZM hard matrix
elements.

However, this seemingly straightforward ``fix" of the conventional
implementation of ZM is not
entirely trivial. The ZM matrix element for the partonic subprocess $%
\gamma ^{\ast }g\rightarrow b\bar{b}$ is obtained after a subtraction,%
\begin{equation}
C^g_{b,\lambda}(\mu )=C^g_{b,\lambda}(1/\varepsilon )_{\mathrm{unsubtracted}%
}-\mathrm{\overline{MS}\,subtraction}(1/\varepsilon ,\mu ),  \label{ZmGF}
\end{equation}%
where the subtraction term represents the collinear singularity due to the
zero quark mass approximation. This term is sensitive to $\mu $, and in
general its $\mu $ dependence matches that of the PDF $b(x,\mu )$ in the LO $%
\gamma ^{\ast }b\rightarrow b$ term. If this LO term is absent because of
the conventional scale choice $\mu =Q$ while $Q<m_{b}$, there would be a
mismatch; then the simple addition of the $\gamma ^{\ast }g\rightarrow b\bar{%
b}$ subprocess, using Eq.\,(\ref{ZmGF}), would be incorrect. However, this
problem arises only because of the choice of the default scale $\mu =Q$, not
because of the ZM formalism itself. For heavy-flavor production, it is in
any case natural to choose a factorization scale $\mu $ $>m_{b}$, e.g.\,$\mu
=\sqrt{Q^{2}+m_{b}^{2}}$. Then the corresponding LO will be present, and
consistency is restored. Clearly, any choice of $\mu$ that stays above $%
m_{b} $ and approaches $Q$ at high energies would be equally acceptable. The
differences will be of higher order. We have verified that the results are
not sensitive to the scale choice, hence will use the above as the default.

There are also other ways of compensating for the
$\overline{\mathrm{MS}}$ subtraction to the gluon-fusion contribution
if the LO processes $\gamma ^{\ast }b\rightarrow b$ and $\gamma ^{\ast
}\bar{b}\rightarrow \bar{b}$ are absent. For instance, in the GM
calculation, the term that removes the collinear
singularity of the unsubtracted gluon fusion term is \cite{ACOT},%
\begin{equation}
\frac{\alpha _{s}}{2\pi }\ln \left( \frac{\mu ^{2}}{m_{b}^{2}}\right) \left[
C_{b,\lambda }^{b(0)}\otimes P_{qg}\otimes f_{g}\right]   \label{colinsub}
\end{equation}%
where $f_{g}$ is the gluon distribution, $P_{qg}$ is the $g\rightarrow
q$ splitting function, and $C_{b,\lambda }^{b(0)}$ is the LO ZM matrix
element for $\gamma ^{\ast }b\rightarrow b$ contribution
to the structure function $%
F_{\lambda }$. This logarithmic term is equivalent to the 
subtraction term in Eq.\,(\ref{ZmGF}), cf.\,\cite%
{ACOT}. Therefore, one can consider adding (\ref{colinsub}) to the ZM
matrix element $C_{b,\lambda }^{g}(\mu )$ in order to maintain
consistency
(scale independence).\footnote{%
Since $f_{b}\approx \frac{\alpha _{s}}{2\pi }\ln (\frac{\mu ^{2}}{m_{b}^{2}}%
)[P_{qg}\otimes f_{g}]$ in the $\mu \sim m_{b}$ region, this term is
essentially the same as the (absent) LO term $[C_{b,\lambda }^{b(0)}\otimes
f_{b}]$. Thus, the underlying physics is similar to that of the default of
the previous paragraph, and the effect on the results turns out to be
similar, too, cf.\,Sec.\,\ref{sec:PdfComp}.}

The modified formulations of the ZM described in the above two
subsections are designed to remedy the obvious problems of the
conventional ZM in both the (large $x,$ small $W$) and the (small $x,$
small $Q$) regions, as identified in Sec.\,\ref{sec:conven}. They
illustrate the point that the simplicity of the ZM parton formalism can
be retained without the manifest violation of physical requirements.
The proposed modifications of the conventional implementation are not
unique. The differences between the modified choices all vanish at
energy scales much larger than $m_{h}$; and they usually are one order
higher in $\alpha _{s}$.

These improvements are all motivated by, and adapted from, recent
implementations of the more accurate GM formalism \cite{cteq65,ThorneTung}.
They belong to an intermediate stage between the ZM and the GM formalisms,
hence will be designated as \emph{intermediate-mass 
variable flavor number schemes%
} (IM VFNS, or IM scheme). They can be viewed either as improved ZM
calculations with physical treatment of final states inspired by GM, or
equivalently, as simplified GM calculations with suitably defined ZM hard
cross sections.

\subsection{Comparison of calculational schemes with heavy-quark data \label{sec:comp}}

To carry out a comparative study, we have implemented the conventional ZM
and the various formulations of the IM scheme described above in our CTEQ
global analysis program, alongside with the existing implementation of the
GM scheme that by now has become the standard.

As a first step, we compare the heavy-quark calculational schemes by
evaluating inclusive DIS cross sections with heavy quarks in the final
state, using known PDFs. These cross sections are most directly affected by
the heavy-quark mass treatment, hence give us a clear measure of the
significance of the various calculational schemes. The data sets consist of
charm production from H1 \cite{H1F2c,H1bcHQ,H1bcLQ}, ZEUS \cite{ZNF2c,ZNF2c1}%
, CCFR and NuTeV \cite{CcfrNutev}; and bottom production from H1
\cite{H1bcHQ,H1bcLQ}. The total number of data points is 222.

For this initial comparison, the cross sections are calculated in the
conventional ZM formulation, the GM formulation, and the IM
formulation of Sec.\,\ref{sec:FS}. 
For each of these calculations, we first evaluate the
cross sections using the PDFs of the latest GM global fit CTEQ6.6M \cite%
{cteq66} (abbreviated as CT66). These numbers are useful as references for
assessing the results. The other PDF set used for comparison is CTEQ6.1M
\cite{cteq61} (abbreviated as CT61), which represents the latest CTEQ ZM PDF
set (2003).\footnote{%
The original CT61 analysis did not include the latest data sets for
semi-inclusive DIS heavy-quark production. The effect of including the
newest heavy-quark data on the ZM PDFs will be described in the
comparisons presented below. \label{fn:ct61}} The $\chi ^{2}$ values
for each combination of the calculational scheme and PDF set serve as a
measure of the goodness-of-fit in the comparison of theory with data.
They are presented graphically in Fig.\,\ref{fig:HqChi}. In the 
two IM calculations, we adjust the parameter $\lambda$ 
in the rescaling variable $\zeta$ of
Eq.~(\ref{zeta})  so as to obtain the lowest possible $\chi^2$ with 
the heavy-quark data for each examined PDF set. \medskip

\noindent * The GM calculation using CT66 (leftmost bar) agrees
perfectly with the experimental data---$\chi^{2}$ of 186 for 222 data
points. The dashed line, corresponding to $\chi^{2}/\mbox{point}=1$, is
drawn for reference.\medskip

\begin{floatingfigure}[r]{3in}
\hspace{-2em}
\includegraphics[width=3in]{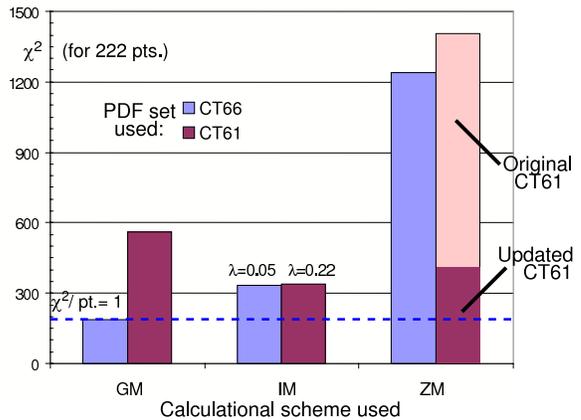}
\caption{The log-likelihood $\chi^{2}$ for 222 data points from
neutral- and charged-current heavy quark semi-inclusive DIS, obtained
in the GM, IM, and ZM schemes.} \label{fig:HqChi}
\end{floatingfigure}
\noindent * The ZM calculation using CT66 PDFs (second to rightmost
bar)
 fails dramatically compared
to data by comparison. Since the CT66 PDFs, as far as we know, are the
closest to the true PDFs, the ZM-CT66 column gives a reasonable measure
of the size of the error of the ZM calculation in the conventional
imple\-men\-ta\-tion---$\chi^{2}\sim1200/222$ points---a very large
discrepancy. The same point is illustrated by the GM-CT61 column
(second leftmost). In this case, since the GM scheme comes closest to
the correct calculation, the discrepancy provides an approximate
measure of how the CT61 PDFs are distorted to compensate for the
imperfect calculation used in the original ZM fit.\medskip

\noindent * The $\chi ^{2}$ value is the largest in the case of the ZM-CT61
calculation. This discrepancy is partially accounted for by the fact that
the heavy-quark data did not exist at the time of the CT61 global analysis.
This number comes down to $\chi^2=406$ if the recent heavy-quark data are included (by
performing an updated ZM analysis), as shown by the darker-colored part of
the ZM-CT61 bar. But the updated CT61 PDFs still do not achieve the level of
agreement with the data observed in the GM-CT66 calculation, as can be
observed from the comparison of the leftmost (GM-CT66) and rightmost
(ZM-CT61) bars in the figure.\medskip

\noindent * The impact of the IM scheme is illustrated by the 
IM-CT66 calculation. We chose  $\lambda=0.05$ in the rescaling
variable $\zeta$ so as to attain the lowest $\chi^2$ with the examined
heavy-quark data sample. The IM scheme
dramatically reduced the ZM-CT66 mismatch, with $\chi
^{2}$ decreased from $1240$ to $337$. Corrective measures adopted in the IM
formulation, addressing manifest flaws of the conventional ZM treatment, do
have a decisive, positive effect. The IM-CT61 fit 
similarly performs better than ZM-CT61, resulting
in the reduction of $\chi^2$ to $347$, reached for $\lambda=0.22$.\footnote{The original ACOT rescaling variable 
$\chi$ (defined in Eq.~(\ref{acotchi}) and equivalent to the  
variable $\zeta$ with $\lambda=0$) produces worse IM fits, with
the log-likelihood $\chi^2=354$ in the CT66 case and $650$ in the  CT61 case.} \medskip

\noindent These results provide an incentive to pursue a more detailed study
of the IM formulation, including a new global fit of PDF's in this
formulation. Since these calculations retain the zero-mass matrix
elements, they are still only approximations to the GM scheme. What we have
seen here is that, in practice, the numerical effects due to the final-state
counting and phase space treatment (that have been corrected) are larger
than those due to the approximate hard matrix elements (that remain). 
Thus, it is possible that the core feature of the ZM scheme---its simple hard 
matrix elements---can be preserved in a viable global analysis.


\section{Global Analyses\label{sec:GlbFits}}

We now turn to the comparison of global analyses performed using the various
schemes. The goal is to quantitatively study how the proposed IM schemes
bridge the gap between the GM scheme (CT66) and the conventional ZM (CT61
and its updated equivalent). We will explore in detail the efficacy of
several realizations of IM scheme described in Sec.\,\ref{sec:alt}, by
comparing both the parton distributions and the physical predictions of
complete global analyses based on these realizations with those of the
conventional ZM and the GM formulations.

\subsection{Comparison of global fits in the ZM, IM and GM schemes}

\begin{floatingfigure}[r]{3in}
\hspace{-2em}
\includegraphics[width=3in]{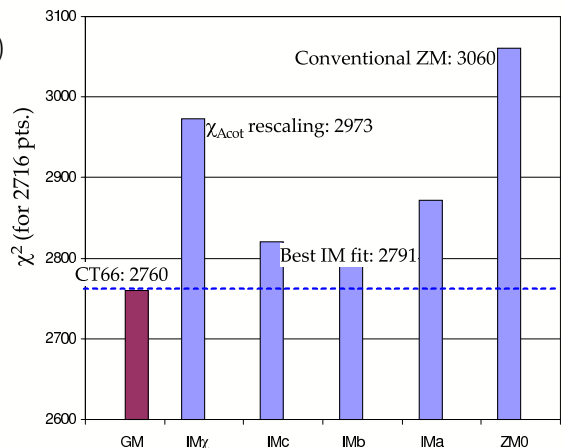}
\caption{The global $\chi^{2}$ values obtained in the GM, IM, and
ZM0 schemes discussed in the text. }\label{fig:GlbChi}
\end{floatingfigure}

We found that among the possibilities explored, the choice of the
rescaling variable (Sec.\,\ref{sec:rescaling}) has the largest impact.
We present here results from four IM global fits, for
$\lambda=0.3$, $0.15$, $0.05$, and $0$ in the rescaling variable $\zeta$
defined in Eq.~(\ref{zeta}). These fits are identified as IMa, IMb,
IMc, and IM$\chi$. The conventional ZM fit (denoted as ZM0) is close
numerically to all IM fits with $\lambda\gtrsim1$. All these fits use
the same experimental input data and parametrizations of the input
nonperturbative PDFs as in the reference CT66 fit (denoted by GM)
\cite{cteq66}, in order to allow meaningful comparisons.

The goodness-of-fit $\chi^{2}$ values in these global fits are shown in
Fig.\,\ref{fig:GlbChi}. For a total of 2716 data points, the best fit is
given by the reference GM fit CT66 ($\chi^{2}=2760$). This is physically
significant---the GM QCD formalism is in better agreement with a wide
variety of global data than all the approximate ZM and IM formulations. The
quality of the IM fits shows sensitive dependence on the value of the $%
\lambda$ parameter that controls the rescaling behavior. The $\chi^{2}$'s of
the two worst fits, ZM0\footnote{%
The ZM0 fit is similar, but not identical, to the earlier CTEQ6.1 fit \cite%
{cteq61}, cf.\,footnote \ref{fn:ct61}. The results of this ZM fit were used
in obtaining the ``updated CT61'' column of Fig.\,\ref{fig:HqChi}.\label%
{fn:zm0}} and IM$\chi$ at the upper and lower ends of this group, are
$\sim300$ higher than in the reference fit (for $2716$ data
points)---beyond the acceptable range of fits characterized by the 90\%
C.L.~criterion of CTEQ analyses ($\sim \Delta\chi^{2}\lesssim100$
\cite{cteq60,cteq61,cteq65,cteq66}). In between these two special
cases, the quality of the global fit spans a
continuous range, with the best fits to the global data occurring for the $%
\lambda$ parameter around $0.15$. The IMb fit ($\lambda=0.15$) has an
overall $\chi^{2}$ of $2791$---only $31$ above that of the CT66
fit. It provides reasonable description of the heavy-quark SIDIS data
examined in Section~\ref{sec:comp}, with $\chi^2=212$ for 222 data points. 
By these criteria, it is well within the usual range of acceptable fits.

All the new IM global fits shown in Fig.\,\ref{fig:GlbChi} use the default
prescription for treating heavy-flavor final states described in Sec.\,\ref%
{sec:FS}. As pointed out there, one can alternatively choose either a
different scale (other than the default $\mu=\sqrt{Q^{2}+m_{h}^{2}}$),
or a different way of compensating the $\overline{\mathrm{MS}}$
subtraction in the ZM gluon-fusion matrix element. We have investigated
the effects due to both of these options on the global fit. They turn
out to be relatively
small compared to those of varying $\lambda$ described above: the overall $%
\chi^{2}$ of the global fit does not change by more than $\sim15$ for all
reasonable alternative choices in either cases (for any given $\lambda$).

Since the IM calculations still retain the ZM hard matrix elements, the IM
PDFs will involve some adjustments made to compensate for this
approximation.  We shall try to quantify the deviations of the PDFs of
the these global fits from those of the GM reference fit, as well as
the possible differences in their predictions in the next section. (We
shall find both to be surprisingly small for the best IM fits.)

\subsection{Comparison of parton distributions\label{sec:PdfComp}}

Poor $\chi^{2}$ values in the bad fits (ZM0 and IM$\chi$) from the previous
section are caused, for the most part, by disagreements with the
high-precision HERA data points at small $x$ and small $Q$. This is
expected, in view of the problems of the ZM scheme and the conventional
rescaling variable $\chi$ at small $x$ discussed in Sections\,\ref{sec:conven}
and \ref{sec:alt}. To gain more insight on this, we show in Figs.\,\ref%
{fig:Pdfa} and \ref{fig:Pdfb} the $u$-quark and the gluon distributions of
the ZM0/IMb/IM$\chi$ analyses, normalized to those of the standard fit
CT66M, at two scales, $Q=2$ GeV and $85$ GeV.
\begin{figure}[p]
\begin{centering}
\includegraphics[width=0.5\columnwidth]{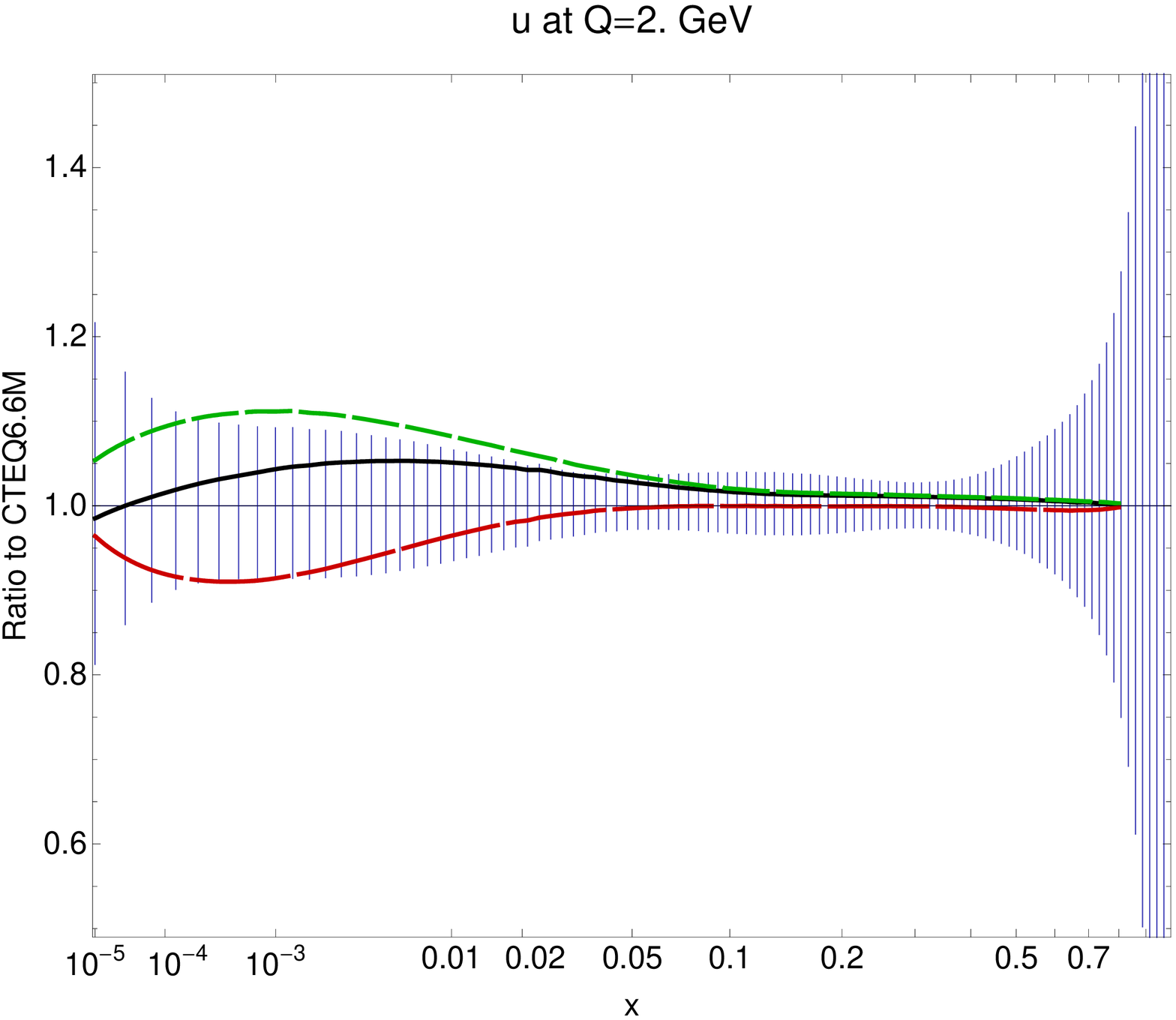}\includegraphics[width=0.5\columnwidth]{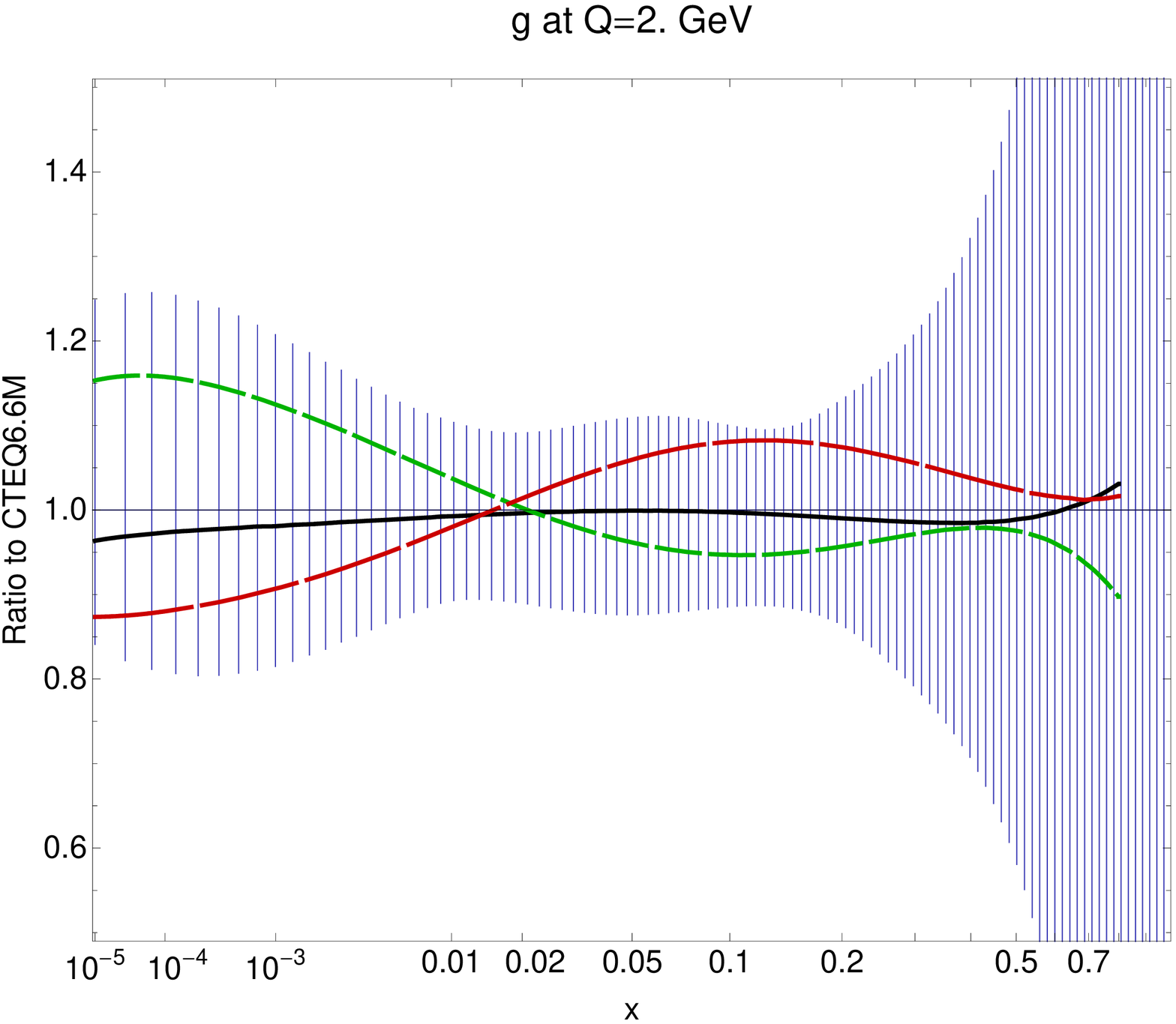}
\par\end{centering}
\caption{The $x$ dependence of the $u$-quark (left) and gluon (right)
PDFs at $Q=2$ GeV, plotted as ratios to the CTEQ6.6M $u$-quark and
gluon PDFs. The blue band is the CTEQ6.6 PDF uncertainty. The lines
correspond to IMb (black solid), ZM0 (red long-dashed), and
IM$\protect\chi$ (green short-dashed) PDFs. } \label{fig:Pdfa}
\end{figure}
\begin{figure}[p]
\begin{centering}
\includegraphics[width=0.5\columnwidth]{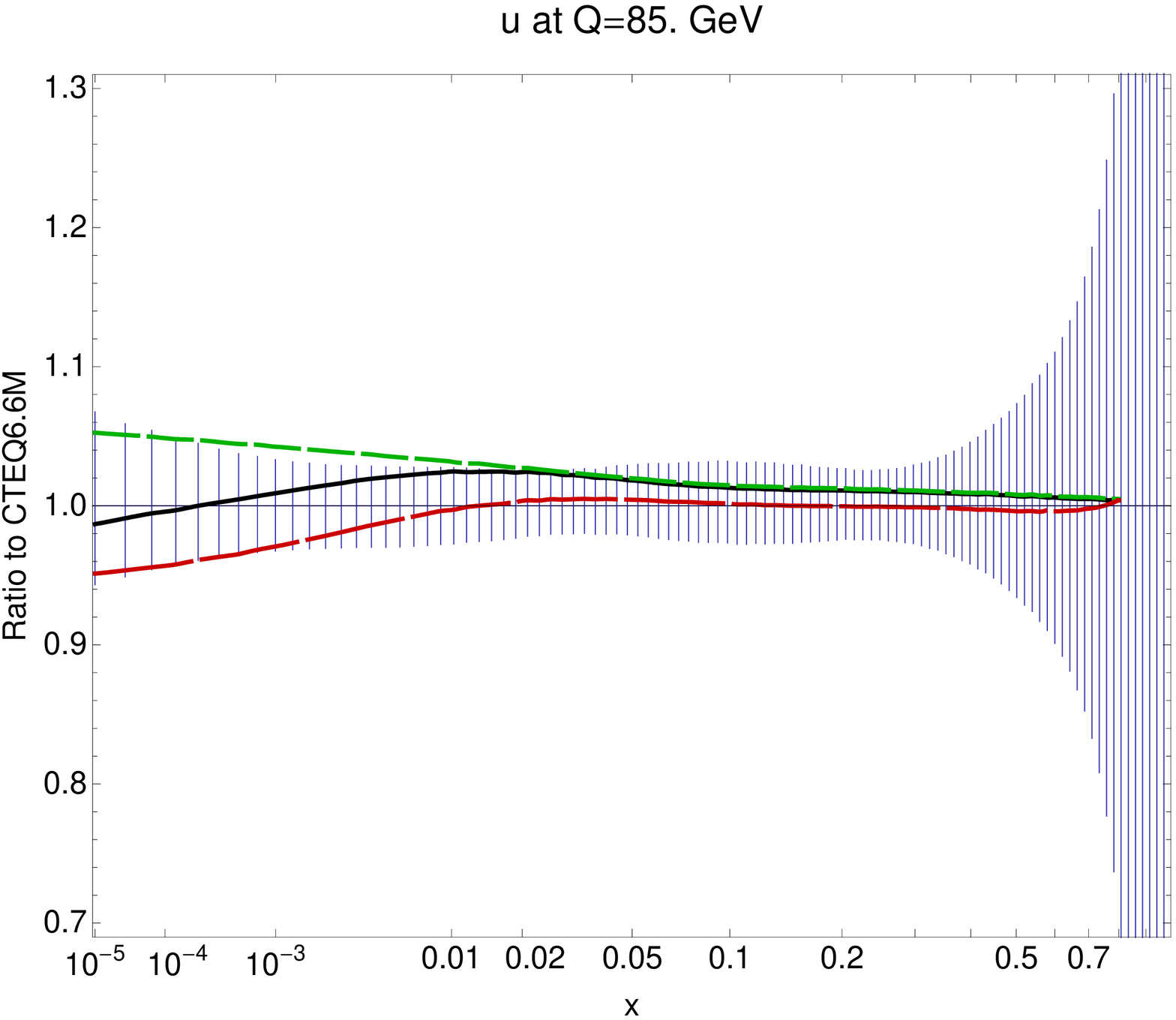}\includegraphics[width=0.5\columnwidth]{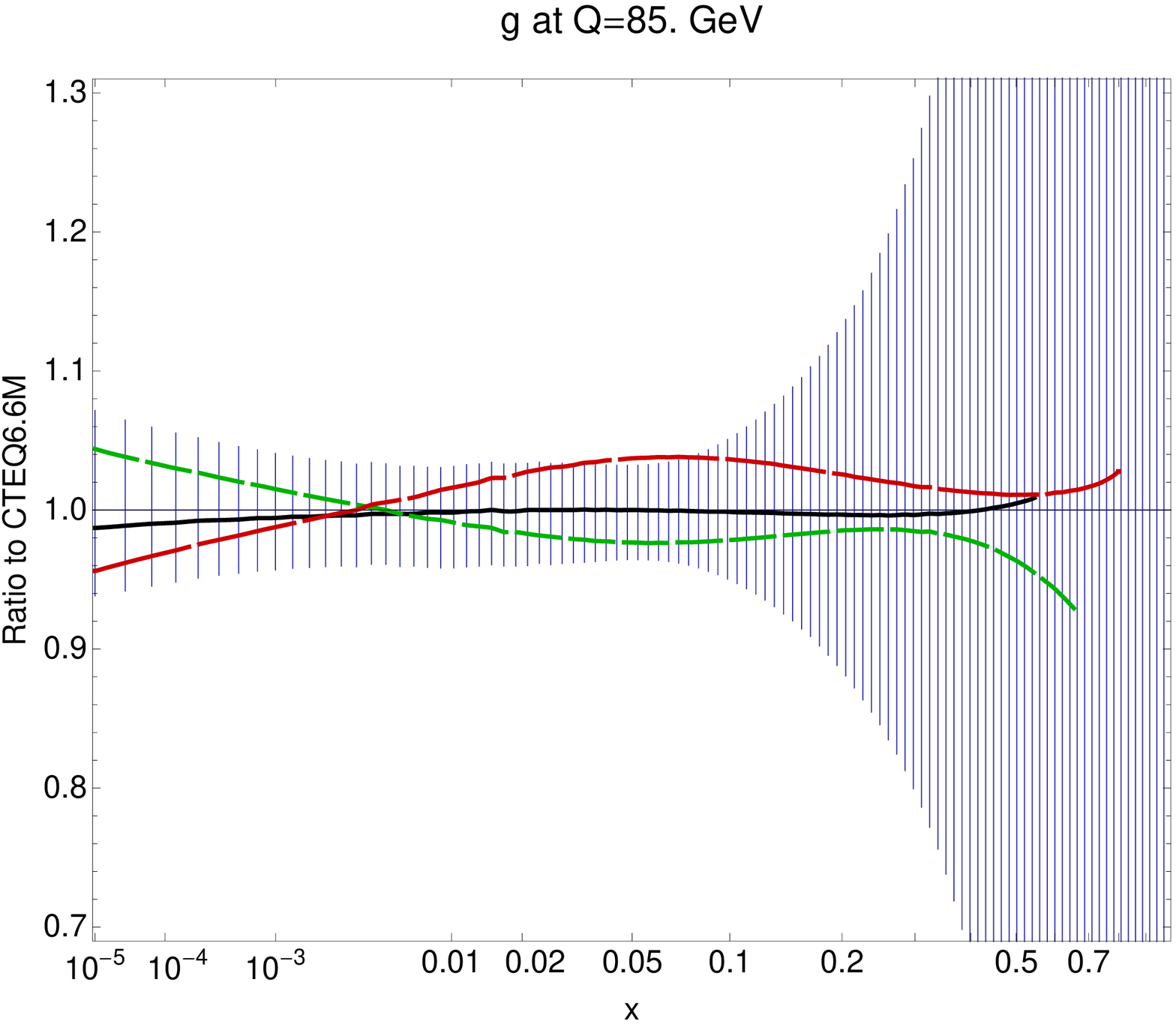}
\par\end{centering}
\caption{Same as Fig.~\protect\ref{fig:Pdfa}, at $Q=85$ GeV. }
\label{fig:Pdfb}
\end{figure}
The $d$-quark distribution
behaves in a similar way as for the $u$-quark (only with larger error
bands), hence is not separately shown. The radiatively generated charm and
bottom PDFs (not shown either) behave similarly to the gluon PDFs. To
provide a useful reference for gauging the size of the deviations, we
also show the uncertainty bands of the CT66 PDFs that result from the
propagation of the input experimental uncertainties.

\begin{figure}[tbh]
\centering{
\includegraphics[width=0.45\columnwidth]{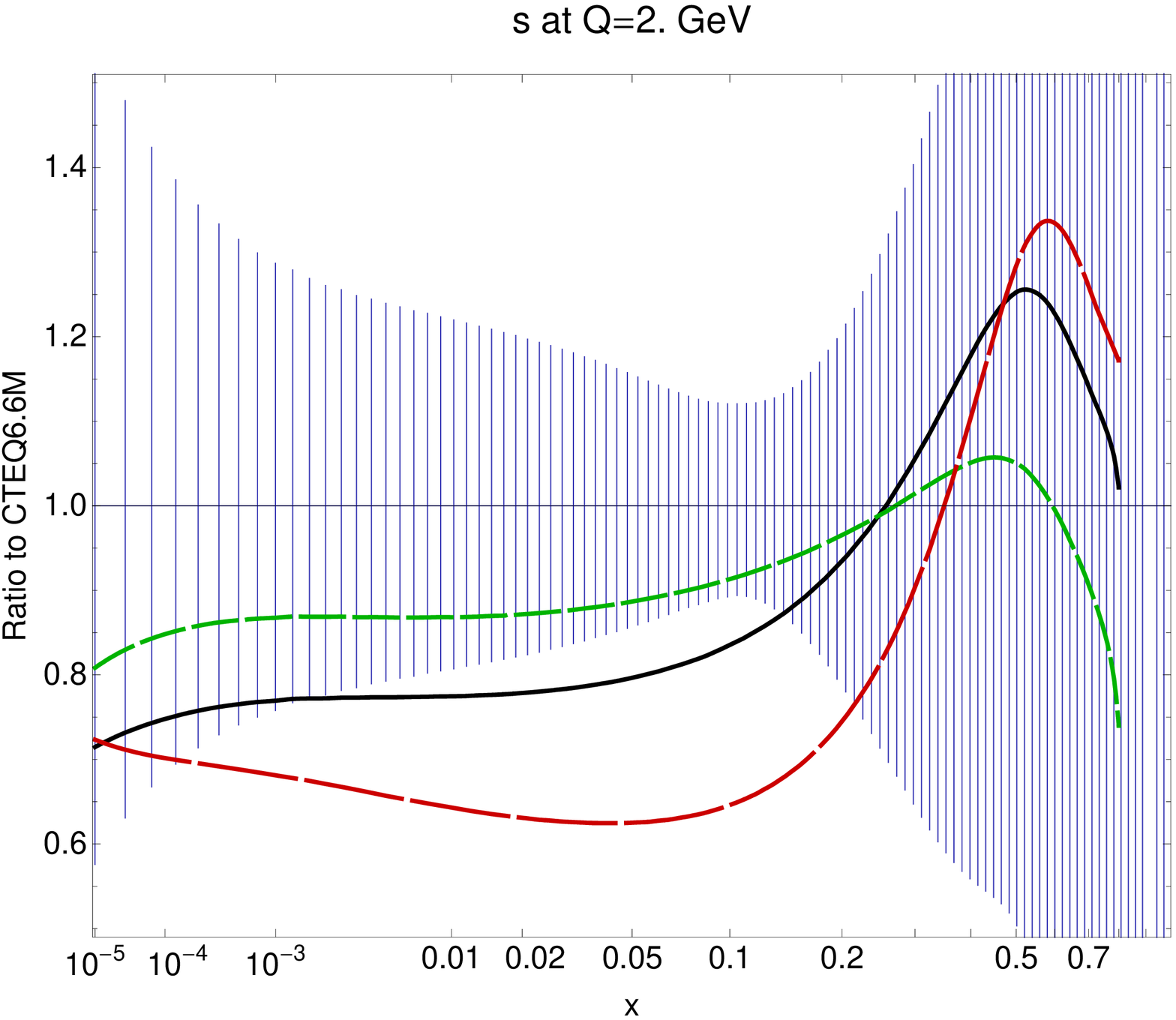}
}
\caption{Same as in Fig.~\protect\ref{fig:Pdfa}, for the strange
quark PDF at $Q=2$ GeV. }
\label{fig:Pdfc}
\end{figure}

We see that the ZM0 and IM$\chi$ PDFs lie on opposite sides of the CT66
PDFs; and the PDFs of the best IM fit, IMb, lie in between. The quark
distributions of the ZM0 set are suppressed compared to those of CT66,
particularly at small $x$. This is because, with no threshold mass
suppression or rescaling in the ZM0 fit, the magnitude of the
heavy-quark scattering contributions to inclusive DIS structure
functions is overestimated (especially at small $x$), resulting in
suppressed light-flavor PDFs in order to compensate for excessive
heavy-quark scattering in the global fit \cite{cteq65,cteq66}. On the
other hand, $\chi$ rescaling introduces too much suppression of phase
space in heavy-quark DIS contributions at small $x$
(cf.\,Sec.\,\ref{sec:rescaling}), resulting in overcompensated
light-flavor PDFs. The gluon distribution shows similar
suppression/enhancement as the quark distribution at small $x$, but
this behavior reverses itself at moderate values of $x$---mostly as a
consequence of the momentum sum rule constraint.

For both ZM0 and IM$\chi$ fits, the deviations of the light quark
distributions from their CTEQ66 counterparts at small $x$ are at the
boundary of the PDF uncertainty bands; whereas those of the $g$, $c$, and $b$
distributions stay generally within the bands. With
the rescaling effect operative at large $x$ (where it is well
motivated), but suppressed at small $x$ (where it is not necessarily
needed on physical ground), the intermediate fit, IMb, yields the PDFs
in-between the two extreme cases. The $u$ and $d$ (anti)quark PDFs in
the IM scheme are consequently closer to the
CT66 reference PDFs (surprisingly so for the gluon). The rescaling variable $%
\zeta$ adopted in Section~\ref{sec:rescaling} is thus capable of bringing
both the global $\chi^{2}$ value and shapes of the majority of the PDFs in
the IM analysis close to those in the full GM scheme.



The strange-quark PDFs $s(x,Q)$, shown in Fig.~\ref{fig:Pdfc}, look more
different than the other flavors. In global analysis, $s(x,Q)$ is not as
well constrained in general because it is only sensitive to data on charm
production in neutrino scattering, which is not very precise. The IM$\chi $
PDF is in better agreement with the CTEQ6.6 PDF than the IMb one, suggesting
that the charged-current cross sections prefer a smaller value of 
$\lambda $ (i.e., closer to ACOT $\chi$) 
than the overall value of 0.15 for the best fit. (The rescaling
variable $\zeta $ has different mass dependence in the two cases.)

\begin{figure}[H]
\centerline{
\includegraphics[width=0.5\columnwidth]{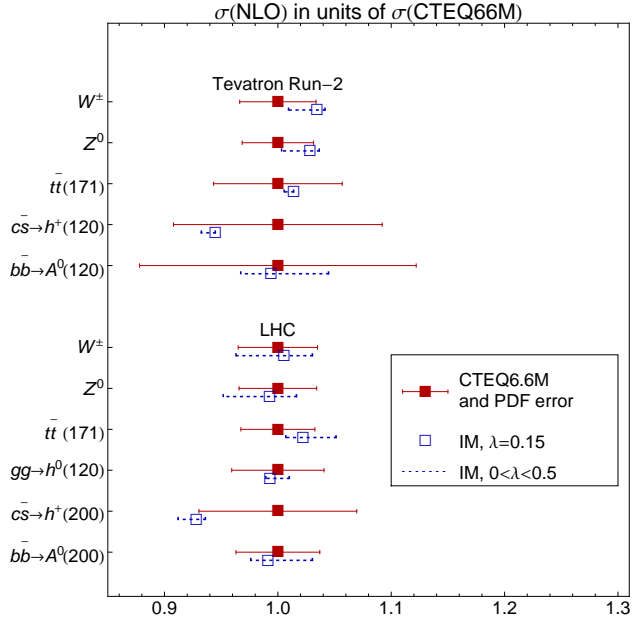}
}
\caption{Representative NLO cross sections at the Tevatron and LHC,
normalized to the CTEQ6.6M cross sections.}
\label{fig:Xsecs}
\end{figure}

\subsection{Comparison of Physical Predictions\label{sec:PhyApp}}

To see how well the IM PDFs do in high-energy physics phenomenology, we
have compared theoretical total cross sections for a number of
representative scattering processes at the Tevatron and the LHC using
the new PDF sets described above. In this calculation, we adopted the
procedure of the analogous comparison in the CTEQ6.6 paper
\cite{cteq66}. To probe typical combinations of parton distributions
and $x$ values, we show cross sections for production of $W^{\pm }$ and
$Z^{0}$ bosons; top quark-antiquark pairs (for top quark mass
$m_{t}=171$ GeV); Standard Model Higgs bosons $h^{0}$ with mass 120 GeV
in $gg$ fusion \cite[and references therein]{Spira:1995mt};
supersymmetric neutral CP-odd Higgs bosons $A^{0}$ with masses 120 and
200 GeV in $b\bar{b}$ annihilation \cite{Spira:1997dg}; and
supersymmetric charged Higgs bosons $h^{+}$ in $c\bar{s}$ scattering
\cite{DiazCruz02}. Details of these calculations are provided in
Section 4 of Ref.~\cite{cteq66}. The cross sections are evaluated at
the next-to-leading-order in the QCD coupling strength $\alpha _{s}$
using the program WTTOT \cite{WTTOT}.


Fig.~\ref{fig:Xsecs} compares the IM and CTEQ6.6 total cross sections.
The central predictions (denoted by boxes) correspond to the best-fit
CTEQ6.6 and IMb PDF sets.  Filled red boxes and error bars denote the
CTEQ6.6 predictions and the PDF errors due to the propagation of
experimental uncertainties. Empty blue boxes are for the best-fit IM
fit with $\lambda =0.15$; the dashed lines represent the range of IM
predictions for the rescaling parameter $\lambda$ in the range $
0\leq\lambda\leq0.5$. We see that the IMb central predictions, and the
IM ranges, mostly lie within the CTEQ6.6 PDF uncertainty bands. The IM
calculational scheme is thus able to reproduce general features of a
variety of GM cross sections. The only exception seen, for
$c\bar{s}\rightarrow h^{+}$ at the LHC, is due to the strange quark as
mentioned at the end of Section~\ref{sec:PdfComp}.

\begin{figure}[tb]
\begin{center}
\includegraphics[width=0.48\textwidth]{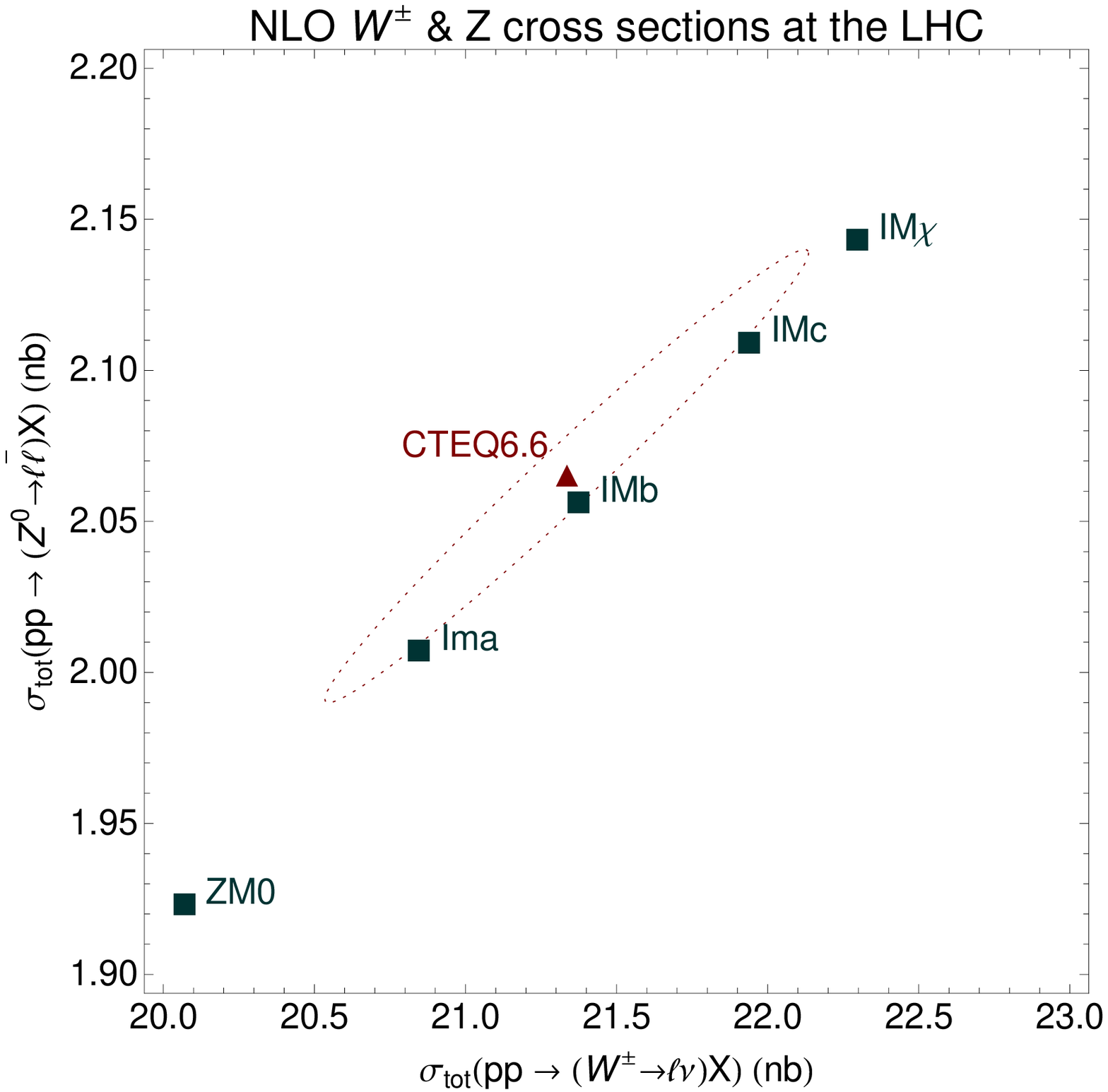} 
\includegraphics[width=0.48\textwidth]{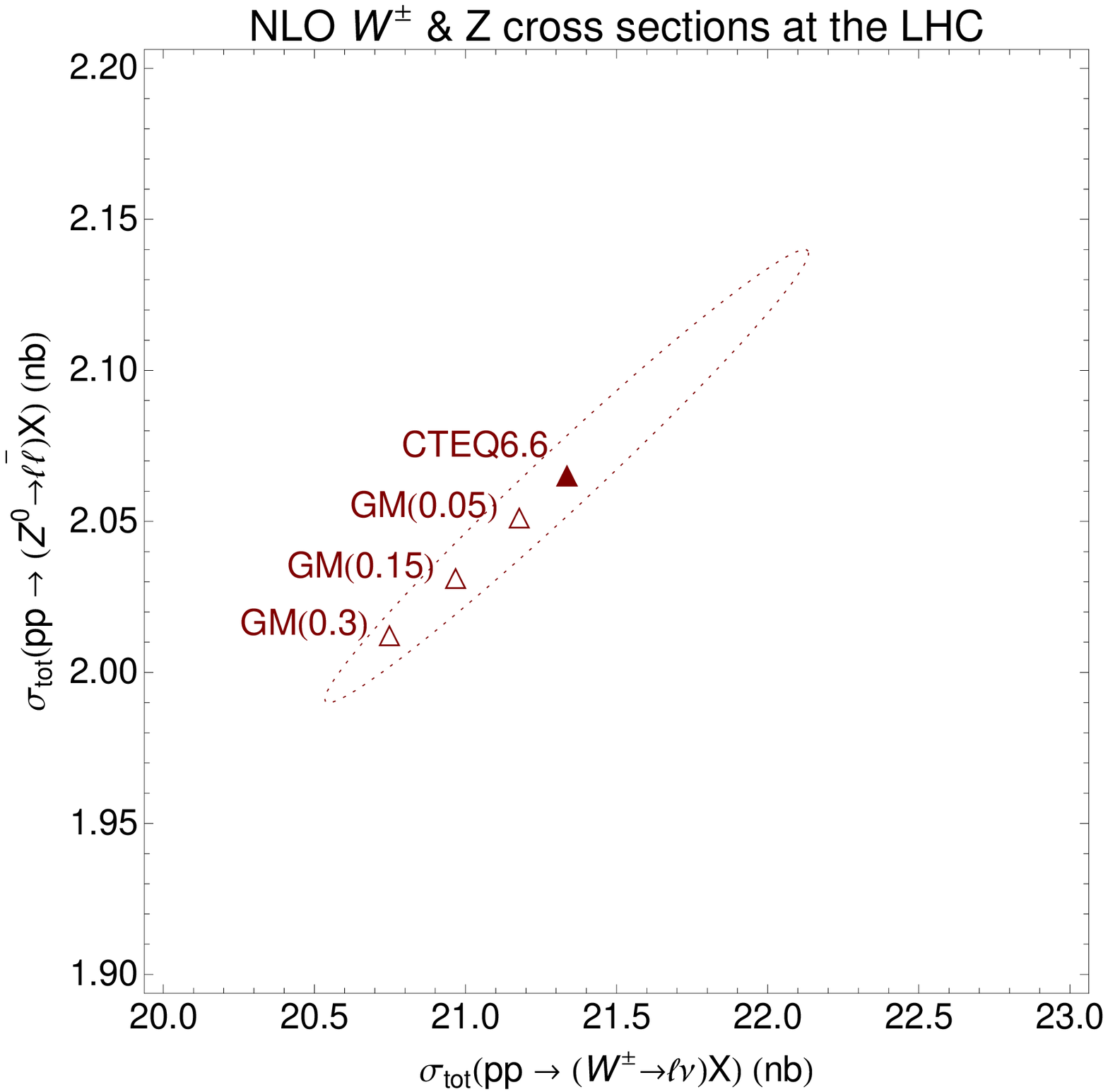}\\ 
(a)\hspace{2.7in}(b)
\end{center}
\caption{NLO
$W^{\pm}$ and $Z$ production cross sections at the LHC, for CTEQ6.6M
and the respective PDF uncertainty (red triangle and dashed ellipse);
for best-fit ZM0, IMa, IMb, IMc, and IM$\chi$ PDFs (left subfigure, filled blue
boxes); for GM fits with $\lambda = 0.05$, $0.15$, $0.3$ (right
subfigure, empty red triangles). 
\label{fig:WZlhc}
}
\end{figure}

The IM results for $W$ and $Z$ production cross sections at the LHC are shown
in the two-di\-men\-sional plot of Fig.\,\ref{fig:WZlhc}(a). The ellipse in the
plot corresponds to the CTEQ6.6 PDF uncertainty range, similar to the bands
shown in the previous figure. We see that the ZM0 and IM$\chi$ predictions are
far away from the CTEQ66M one, in opposite directions; while the
intermediate ones are close to the boundary of the uncertainty ellipse. Not
surprisingly, the IMb point is very close to the reference CT66 one. These
results mirror closely the range of variation of the $u$-quark distribution
at $\sim10^{-3}$ in Fig.\,\ref{fig:Pdfb} (with similar results for the $d$
quark that is not shown), since the $W/Z$ cross sections are dominated by
the quark-antiquark annihilation process with the kinematic variable range $%
x\sim10^{-3}$ and $Q\sim80\mathrm{-}90$ GeV. In particular, we see that the
IMb quark line in Fig.\,\ref{fig:Pdfb}a is extremely close to CTEQ6.6M
(horizontal line) in this range.

The IMb prediction in Fig.~\ref{fig:WZlhc}(a) lies in the lower part of the
CTEQ6.6 error ellipse, suggesting that the IMb ratio of the $Z$ and $W$
total cross section, $\sigma_{tot}(Z)/\sigma_{tot}(W)$, is lower than in the
CTEQ6.6M case. This is again consistent with the observation that the IMb
strange PDF $s(x,Q)$ is lower than the CTEQ6.6 one in the range of $%
x\sim10^{-3}$ typical for $W$ and $Z$ production at the LHC. As noticed in
Ref.~\cite{cteq66}, the ratio $\sigma_{tot}(Z)/\sigma_{tot}(W)\approx0.1$ at
the LHC, despite being well-constrained, is quite sensitive to the form of $%
s(x,Q)$. A smaller magnitude of $s(x,Q)$ thus results in a smaller $%
\sigma_{tot}(Z)/\sigma_{tot}(W)$ ratio.

\subsection{Generalized rescaling variable in the GM
  scheme\label{sub:zetaGM}}

The IM scheme results discussed above are sensitive 
to the choice of the parameter $\lambda $ in the definition of 
$\zeta$--- an indication of
its phenomenological origin. The choice of the scaling variable, involving
powers of $m/Q$, is inherently an issue beyond the usual perturbative
formalism. In fact, this ambiguity exists also in the GM scheme.

In principle, we could apply the proposed generalized rescaling
variable $\zeta$ to the GM analysis (in place of the default
ACOT-$\chi$) as well. Will the GM formalism be stable with respect to
variations of the $\lambda$ parameter in $\zeta$? We expect the answer
to be ``yes'', since, as a PQCD theory including heavy-quark masses as
the basic parameters of the Lagrangian, the GM formalism contains built-in
compensation between the $\zeta $ dependence of the contributions from
heavy-quark-initiated subprocesses and corresponding subtraction terms
applied to gluon-fusion subprocesses at each order of $\alpha_s$.

We have verified that this expectation indeed holds in practice.
Specifically, in contrast to the large difference in the overall
$\chi^{2}$
of about 180 between the global fits IM$\chi$ ($\lambda=0$) and IMb ($%
\lambda=0.15$), the difference between the $\chi^{2}$ values of two GM
fits for these same $\lambda$ values is only $12$. The CTEQ6.6M PDF,
corresponding to $\lambda=0,$ has a slightly better $\chi^{2}$ than the
GM fits with non-zero $\lambda$ values, suggesting that the ACOT
variable $\chi$ (motivated by the exact kinematics of the heavy-quark
pair production in the gluon fusion process) does the best job of all
in the GM case. The GM PDFs for $\lambda$ as high as 0.4 remain within
the band of the CTEQ6.6 PDF uncertainty, while the IM fit with $\lambda=0.4$
is strongly disfavored. Predictions for the LHC $W$ and $Z$ cross
sections based on GM fits with $\lambda < 0.3$ lie
within the CTEQ6.6 PDF error ellipse, as illustrated by Fig.~\,\ref{fig:WZlhc}(b).
Sensitivity of GM cross sections to $\lambda$ is clearly reduced 
in comparison to the IM calculation. 


\section{Concluding Remarks \label{sec:Conclusions}}

For many years, the ZM variable-flavor number factorization scheme has
given a high-quality description of existing global hard-scattering
data and provided predictions for a wide range of high-energy
processes. Even though by now the GM scheme has superseded the ZM
scheme as the more precise formalism, the ZM scheme still has a lot of
appeal for practical reasons. It continues to be use in most
phenomenological calculations. For this reason, we carry out the present
study to lay out more explicitly the approximations and inconsistencies
inherent in the conventional implementation of the ZM scheme. We show
how these inconsistencies, due to the \emph{ad hoc} imposition of the
heavy-quark mass thresholds on the zero-mass QCD theory, can be
corrected by a more careful physical treatment of the heavy-flavor
final states, while preserving the simplicity of the ZM
hard matrix elements. Our proposed intermediate-mass (IM) calculational
scheme can be considered either as \emph{improved ZM formulations} with
GM kinematics of final states, or \emph{simplified GM formulations}
with ZM hard matrix elements.

The key element that makes the IM scheme useful is the introduction of
a flexible rescaling variable $\zeta$ that generalizes the
mass-dependent rescaling variable $\chi$ \cite{acotchi} that has
already been adopted by the latest global analyses in the GM scheme.
The $\zeta$ variable effectively implements kinematic mass threshold
constraints in heavy-quark production processes, while minimizing
unintended effects away from the physical threshold region in a smooth,
controlled way by a parameter $\lambda$,
cf.\,Sec.\,\ref{sec:rescaling}. This allows us to systematically
investigate how the conventional ZM scheme can be improved while
keeping the simple, well-known, ZM matrix elements.

We demonstrated that global analysis carried out in the IM scheme can
approximate the GM scheme results quite well, both in terms of the
resulting PDFs, and in terms of typical physics predictions at the
Tevatron and the LHC. The IM scheme can play a useful role in
bringing the existing NLO analyses based on ZM hard matrix elements closer
to the GM formulation, even without the full implementation of
heavy-quark mass effects. Dependence of the IM predictions on 
the form of the  effective rescaling variable underlines the phenomenological
nature of this approach. Although this dependence in principle also
arises in the GM formalism, we have demonstrated that it is less
pronounced than in the phenomenological IM formulation. 
Thus, this additional source of theoretical uncertainty 
hardly affects what we know about the GM formalism -- except that, 
perhaps, it should be added to the other sources
of theoretical errors, such as scale dependence, 
when assessing the uncertainty of the GM
theoretical results.

\bigskip

\paragraph{Acknowledgments}

WKT would like to thank Robert Thorne for useful discussions related to
this subject during collaborative work on Ref.\,\cite{ThorneTung}, and
for cogent remarks on an early draft of this paper. 
We thank Paul Thompson for a critical reading of the manuscript 
and many helpful suggestions for improvements of the presentation.
This work is supported by the
National Science Foundation (USA) under the grant PHY-0354838. PMN is
partly supported by the U.S.\ Department of Energy under grant
DE-FG02-04ER41299, and by Lightner-Sams Foundation.

\end{document}